\documentclass[twocolumn,longbibliography,prl,tightenlines,superscriptaddress]{revtex4-2}

\usepackage{hyperref}
\usepackage{amsmath,amssymb}
\usepackage[utf8]{inputenc} 			
\usepackage[T1]{fontenc}				
\usepackage{graphicx}
\usepackage{float}
\usepackage[usenames,dvipsnames]{xcolor}
\usepackage{tikz}

\newcommand{\br}{\textbf{r}}

\graphicspath{{./Image/}}

\newcommand\bi{{\bf i}}

\usepackage{lipsum}
\newcommand\bfr{{\bf r}}

\usepackage{tikz}

\newcommand\cf{c_{\rm f}}

\newcommand\xf{{\hat x}_{\rm f}}
\newcommand{\ro}{{\rm o}}
\newcommand{\md}{m_{\rm d}}

\begin{document}

\title{Metastability of Discrete-Symmetry Flocks}

\author{Brieuc Benvegnen}
\affiliation{Sorbonne Universit\'e, CNRS, Laboratoire de Physique Th\'eorique de la Mati\`ere Condens\'ee, 75005 Paris, France}

\author{Omer Granek}
\affiliation{Department of Physics, Technion – Israel Institute of Technology, Haifa 32000, Israel}

\author{Sunghan Ro}
\affiliation{Department of Physics, Massachusetts Institute of Technology, Cambridge, Massachusetts 02139, USA}

\author{Ran Yaacoby}
\affiliation{Department of Physics, Technion – Israel Institute of Technology, Haifa 32000, Israel}

\author{Hugues Chat\'{e}}
\affiliation{Service de Physique de l'Etat Condens\'e, CEA, CNRS Universit\'e Paris-Saclay, CEA-Saclay, 91191 Gif-sur-Yvette, France}
\affiliation{Computational Science Research Center, Beijing 100094, China}
\affiliation{Sorbonne Universit\'e, CNRS, Laboratoire de Physique Th\'eorique de la Mati\`ere Condens\'ee, 75005 Paris, France}

\author{Yariv Kafri}
\affiliation{Department of Physics, Technion – Israel Institute of Technology, Haifa 32000, Israel}

\author{David Mukamel}
\affiliation{Department of Physics of Complex Systems, Weizmann Institute of Science, Rehovot
  7610001, Israel}

\author{Alexandre Solon}
\affiliation{Sorbonne Universit\'e, CNRS, Laboratoire de Physique Th\'eorique de la Mati\`ere Condens\'ee, 75005 Paris, France}

\author{Julien Tailleur}
\affiliation{Department of Physics, Massachusetts Institute of Technology, Cambridge, Massachusetts 02139, USA}

\date{\today}

\begin{abstract}
We study the stability of the ordered phase of flocking models with a scalar order parameter.
Using both the active Ising model and a hydrodynamic description, we show that droplets of particles moving in the direction opposite to that of the ordered phase nucleate and grow. We characterize analytically this self-similar growth and demonstrate that droplets spread ballistically in all directions. Our results imply that, in the thermodynamic limit, discrete-symmetry flocks 
---and, by extension, continuous-symmetry flocks with rotational anisotropy---
are metastable in all dimensions.
\end{abstract}

\maketitle

Spin waves are well known to prevent the emergence of true long-range order
in the two-dimensional (2d) equilibrium $XY$ model, as dictated by the Mermin-Wagner theorem~\cite{mermin1966absence,kardar2007statistical}. It was therefore a surprise when the Vicsek model~\cite{vicsek1995novel}, which can be seen as an active version of the $XY$ model, was shown to possess a flocking phase with true long-range polar order, `escaping' the theorem. 
In this `Toner-Tu' phase~\cite{toner1995long}, the model exhibits giant (scale-free) density and order parameter fluctuations~\cite{toner1995long,toner2005hydrodynamics,toner2012reanalysis,chate2008collective,mahault2019quantitative}.
Since long-range order survives spin waves,
it seems reasonable that, as in equilibrium, the ordered phase should also survive stronger excitations. 
This may be part of the reason why its robustness to non-perturbative fluctuations
has long remained unexplored.

Recently, however, numerical results in 2d have questioned the
stability of the flocking phase.  First, it has been shown that, for
large enough noise, a small obstacle or a group of particles oriented
against the flow can break order in the Vicsek model by triggering a
counter-propagating front that grows and eventually reverses the
global polarity~\cite{codina2022small}.  This happens in a finite
fraction of the phase diagram close to the disordered phase, hence
shifting the flocking phase to lower noise values.  Second,
constant-density flocks have been shown to be metastable to the
spontaneous nucleation of aster-shaped
defects~\cite{besse2022metastability}.  These numerical results
question the stability of the 2d orientationally-ordered phases of
active systems. So far this question has not been approached theoretically. Furthermore, higher dimensions remain uncharted territory.

\begin{figure}[th!]
    \centering
    \includegraphics[width=\columnwidth]{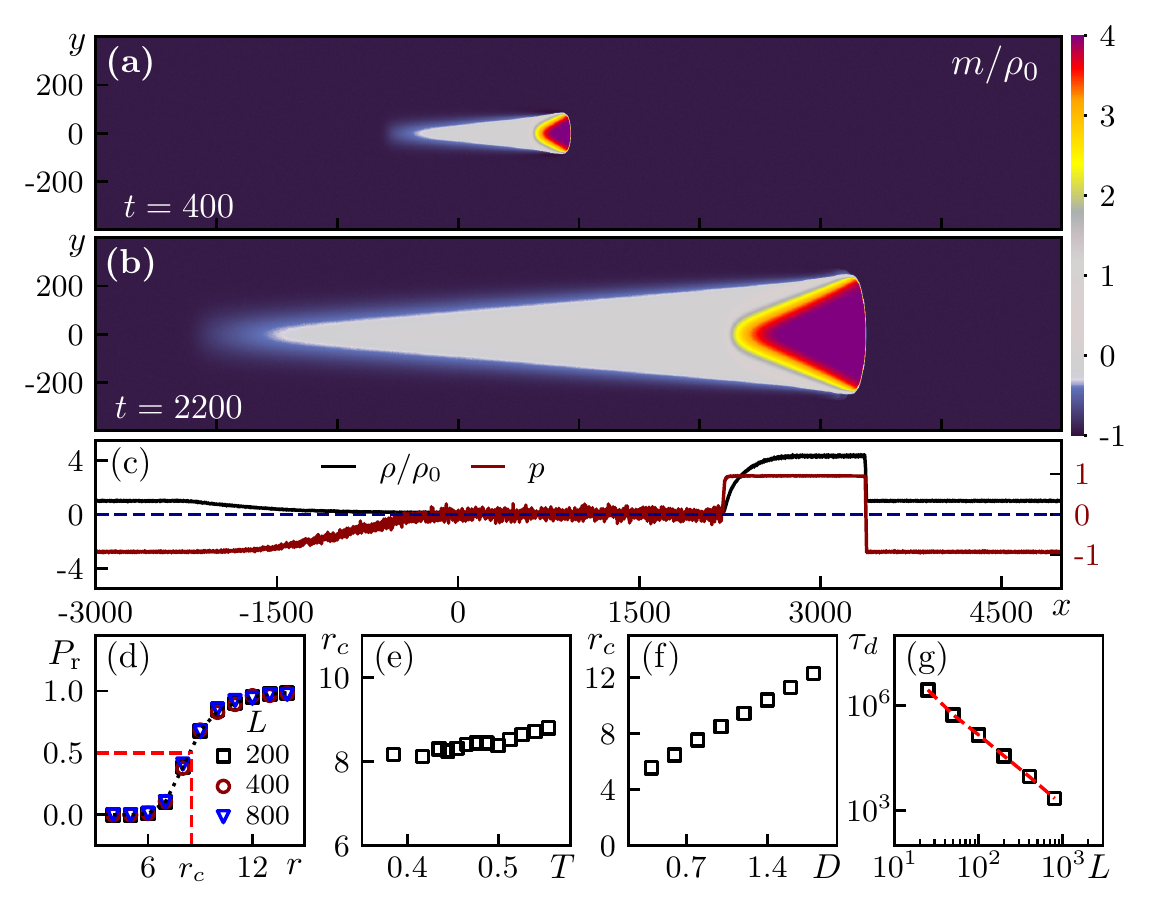}
    \caption{Simulations of the 2d AIM. 
    {\bf (a-b)} Snapshots of the magnetization field following the introduction in the ordered phase 
    of a counter-propagating droplet at $x=y=t=0$ ($r=10$, $\rho_{\rm d}^0=5\rho_{\rm o}$), averaged over 100 runs. 
    {\bf (c)} Polarization (red) and density (black) profiles at $y=0$ for the snapshot shown in (b). 
    {\bf (d)} Reversal probability $P_\mathrm{r}(r)$ for $\rho_{\rm d}^0=1.2 \rho_{\rm o}$. 
    The dotted line is a fit to a hyperbolic tangent
    used to extract $r_{\rm c}$. 
    {\bf (e,f)} Variations of $r_{\rm c}$ with $T$ and $D$. 
    {\bf (g)} Average nucleation time $\tau_{\rm d}$ in a system of size $L\times L$. 
    Parameters: 
    $v=1$; $D=1$ (except for (g) where $D=0.1$); $\rho_0=10$ (a-c), $5$ (d-f), and $8$ (g); 
    $\beta = 2$ (except for (f));  $(L_x, L_y) = (8000, 800)$ in (a-c) and $(200, 100)$ in (e-f).
}
    \label{fig:droplet-micro}
\end{figure}

In another recent development, the Toner-Tu phase was shown to be fragile to rotational anisotropy~\cite{solon2022susceptibility}. A field with $q$-fold symmetry influencing particle orientations  
destroys the scale-free nature of the ordered phase at large scales, leading to Gaussian fluctuations and short-range correlations. The ordered phase then displays a behavior akin to that of the active Ising model (AIM), 
the active counterpart to the Ising model~\cite{solon2013revisiting}. 
Thanks to its discrete-symmetry, the AIM is more amenable to analytical treatments, 
which made it instrumental for establishing the transition to collective motion as a non-equilibrium liquid-gas
phase separation~\cite{solon2013revisiting}. (Except in $d=1$ where the ordered phase is, as in equilibrium, unstable~\cite{o1999alternating,solon2013revisiting,benvegnen2022flocking}).
Equilibrium physics suggests that the discrete symmetry of the AIM should make its ordered phase more robust to fluctuations than that of the Vicsek model. Indeed
the stability of the ordered phase of the equilibrium Ising model against nucleation
is guaranteed by Peierls' argument in $d>1$~\cite{griffiths1972phase,peierls1991more},
which essentially states that surface tension makes any finite droplet of the minority phase shrink.

In this Letter, we show that a very different scenario takes place: the ordered phases of flocking models with a scalar order parameter are metastable in any $d\geq 2$. We first consider numerical simulations of the AIM in 2d and demonstrate that a large enough ---but finite--- 
droplet polarized oppositely to the 
surrounding ordered background grows ballistically in all directions
(see Fig.~\ref{fig:droplet-micro}(a-c) for an illustration in 2d).
We then show numerically and analytically that the same scenario takes place in the relevant hydrodynamic theory. 
We find that droplet growth is induced by the ballistic propagation of its domain walls and characterize its shape in any dimension $d\geq 2$. 
In the asymptotic large size and time limit, multiple droplets spontaneously nucleate, as confirmed by our 2d numerics, hence destroying long-range order.

{\it Active Ising model.}
In the AIM~\cite{solon2013revisiting,solon2015flocking,sakaguchi2019flip,yu2022energy,mishra2022active,kourbane2018exact,chate2020dry}, particles carrying a spin $s=\pm 1$ move on a $d$-dimensional hypercubic lattice. 
The self-propulsion of particles takes place along the lattice vector ${\bf e}_x$,
in a direction given by their spins: particles at site $\bi$ hop at rate $v$ to the site $\bi+s {\bf e}_x$. 
In addition they also hop with a rate $D$ to any nearest-neighbor site.
Occupation is not restricted and we denote by
$n_\bi^{\pm}$ the number of particles on site $\bi$ with spin $s=\pm 1$. Particles experience on-site ferromagnetic alignment
by flipping their spins with a rate $\omega_0 \, e^{-\beta s {m_\bi}/{\rho_\bi}}$, where $\rho_\bi=n^+_\bi+n^-_\bi$ and $m_\bi=n^+_\bi-n_\bi^-$ are
the site density and magnetization. 
In the absence of particle hopping, each site hosts an equilibrium fully connected Ising model at temperature $T=1/\beta$.
We choose time and length units by fixing $\omega_0=1$ and the lattice spacing $a=1$. 
Previous work has reported a phase diagram with three phases~\cite{solon2013revisiting,solon2015flocking}: 
a disordered gas phase and an ordered, flocking, liquid phase, separated by 
a coexistence region where a macroscopic dense liquid domain moves in the surrounding gas.

\begin{figure}[t!]
    \centering
    \includegraphics[width=1.\columnwidth]{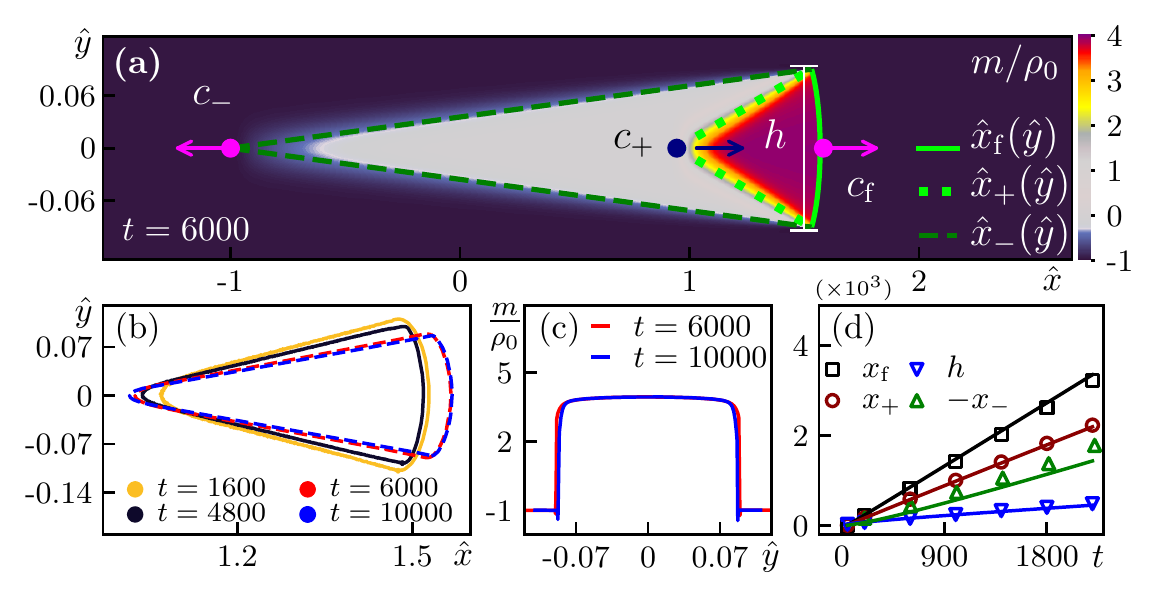}
    \caption{{\bf (a)} Snapshot of the $m/\rho_{\rm o}$ field
    obtained by integrating Eqs.\,\,(\ref{eq:dynrho},\ref{eq:dynm}) starting from a counter-propagating droplet 
     introduced in the ordered phase solution at $t=0$ ($r = 10$, $\rho_{\rm d}^0 = 12\rho_0$). 
     Straight and parabolic green lines are guides to the eye.
    {\bf (b)} Isodensity curves at $\rho = (\rho_{\rm d}+\rho_{\rm o})/2$ in ballistically-rescaled coordinates  
    for microscopic (solid lines) and hydrodynamic (dashed lines) simulations. 
    {\bf (c)} Cross-sectional plots of the magnetization in PDE simulations along the maximal-width line shown in white in (a).
    {\bf (d)} Positions of the front ($x_{\rm f}$) and rear ($x_+$) droplet interfaces, and of the end of the comet ($x_-$)
    at $y=0$, together with the droplet width $h$. Symbols and lines correspond to microscopic and PDE simulations, respectively. 
    Parameters: As in Fig.~\ref{fig:droplet-micro}(a-b) except $(L_x,L_y) = (8000, 600)$ in panel (a).}
    \label{fig:micro-PDE}
\end{figure}

Figure~\ref{fig:droplet-micro} illustrates why and how the ordered phase is in fact only metastable. (Here and in what follows, all our numerical methods are detailed in SM~\cite{supp}.) 
Introducing a circular droplet of radius $r$ and density $\rho_{\rm d}^0=1.2\rho_{\rm o}$ containing counter-propagating particles
in an ordered phase of density $\rho_{\rm o}$, we observe that, for $r$ large enough, 
the droplet (red region) grows at the expense of the ordered phase, leaving in its wake a dilute, disordered comet-like structure (white region) (Fig.~\ref{fig:droplet-micro}(a,b)
and Supplementary Movie~1).
These different regions are separated by domain walls that we characterize later.
For small $r$, the perturbation typically disappears and the ordered phase is restored.
The local density and polarization, defined as $p_\bi\equiv m_\bi/\rho_\bi$, in these three regions are illustrated in Fig.~\ref{fig:droplet-micro}(c), which shows their profile along the central axis $y=0$. 
The maximal density $\rho_{\rm d}>\rho_{\rm o}$ and magnetization $m_{\rm d}$ take well-defined, finite limiting values at late times. 
We note that the polarization inside the droplet is constant and approximately opposite to that of the ordered phase: $p_{\rm d} \simeq -p_{\rm o}$.

Varying $r$ allows us to estimate the
radius $r_{\rm c}$ beyond which the initial droplet grows~\footnote{Keeping
  $r$ fixed and increasing $\rho_{\rm d}^0$ leads to similar results, see
  Fig.\,\,S2}.  Repeating the procedure many times, we find that
 the probability that the droplet reverses the
ordered phase, that we denote by $P_\mathrm{r}(r)$, increases sharply from near-0 to near-1 values when $r$
is increased and we define $r_{\rm c}$ by
$P_\mathrm{r}(r_{\rm c})=\tfrac{1}{2}$. 
We find that $P_\mathrm{r}(r)$ becomes rapidly independent of the system size
 as the latter is increased (Fig.~\ref{fig:droplet-micro}(d)) and measure finite values of $r_{\rm c}$ at
all parameter values probed in the ordered phase.  Remarkably, $r_{\rm c}$
decreases weakly as $T$ is decreased deeper into the `ordered' phase
(Fig.~\ref{fig:droplet-micro}(e)). This is markedly different from
what was reported for the Vicsek model~\cite{codina2022small} where a
perturbation reverses the ordered phase only in a finite fraction of
the phase diagram next to the coexistence region, with $r_{\rm c}$ diverging
sharply as $T$ is decreased.  Note that $r_{\rm c}$ increases roughly
linearly with $D$, consistently with the intuition that diffusion
suppresses perturbations (Fig.~\ref{fig:droplet-micro}(f)).
Finally, we found $r_{\rm c}$ to diverge in the limit $v\to 0$ (see
  Fig.\,\,S2), therefore recovering at $v=0$ an equilibrium-like behavior where finite-size droplets shrink.

The above results indicate that $r_{\rm c}$ can reach relatively small
values at small $D$.  At such parameters, we are actually able to
observe the {\em spontaneous} nucleation of a droplet within the
ordered phase (see Fig.~S3 and Supplementary Movie~2).  Repeated
experiments allow us to estimate $\tau_{\rm d}$, the mean time separating
flow-reversing nucleation events. We find that $\tau_{\rm d}\sim 1/L^2$
(Fig.~\ref{fig:droplet-micro}(g)). This is consistent with nucleation
events resulting from local fluctuations, which, together with the
behavior of $r_{\rm c}$ shown in Fig.~\ref{fig:droplet-micro}, leads us to
conclude that the whole flocking phase is metastable in the
thermodynamic limit.

{\it Hydrodynamic description.} To characterize the mechanisms underpinning droplet growth and to test their robustness to microscopic details, we now study a continuum description of the AIM. Following established procedures used to study coarsening and nucleation~\cite{bray2002theory,krapivsky2010kinetic}, we focus on the low-$T$ large-density regime where the ordered phase is expected to be most stable. We thus consider the mean-field hydrodynamics of the AIM~\cite{solon2015flocking} in $d$ dimensions:
\begin{align}
    \partial_t \rho &= - v \partial_x m + \nabla \cdot \bar D \nabla \rho \label{eq:dynrho} \\
    \partial_t m &= - v \partial_x \rho + \nabla \cdot \bar D \nabla m + F(\rho,m) \label{eq:dynm}\;,
\end{align}
where $\rho$ and $m$ are the density and magnetization fields, respectively, $F=2\rho \sinh(\beta m/\rho)-2m\cosh(\beta m/\rho)$, and $\bar D$ is a diagonal matrix whose elements are $D_{x}=D+v/2$ and, for any other coordinate, $D_{\perp}=D$. Numerical integration of Eqs.\,\,(\ref{eq:dynrho},\ref{eq:dynm}) in 2d, shown in Fig.~\ref{fig:micro-PDE}(a) for the same parameters as in Fig.~\ref{fig:droplet-micro}(a-c), recapitulate the phenomenology observed in microscopic simulations: an initial droplet grows, leaving in its wake a low-density comet. Both particle and field simulations show a late-time dynamics which is self-similar under a ballistic scaling $(\hat x,\hat y)=(x/t,y/t)$ (see Fig.~\ref{fig:micro-PDE}(b) and (c)). To compare them, we identify 
the three interfaces between the ordered background, the droplet, and the comet, which we parametrize as $\hat x_{\rm f}(\hat y)$, $\hat x_+(\hat y)$, and $\hat x_-(\hat y)$, respectively (Fig.~\ref{fig:micro-PDE}(a)).  
Measurements of the unscaled front positions at $y=0$ and of the droplet width show linear growth in time, consistent with ballistic scaling (Fig.~\ref{fig:micro-PDE}(d)). The corresponding speeds measured in microscopic
and PDE simulations agree within a few percents. We now turn to the analytical study of  Eqs.\,\,(\ref{eq:dynrho},\ref{eq:dynm}) in $d\geq 2$.

{\it Propagating domain walls.} Assuming that the axisymmetry observed in 2d is maintained in higher dimensions, we distinguish coordinates parallel and normal to the self-propulsion: $\br=x {\bf e}_x + \br_\perp$. We first characterize the domain walls connecting the droplet, the comet, and the ordered background along the symmetry axis of the droplet ($\br_\perp=0$) before discussing the full droplet shape.
Due to the ballistic scaling, we expect that the unscaled interfaces become locally flat at $\br_\perp=0$ at late times in any direction normal to ${\bf e}_x$. The domain walls connecting the different regions at $\br_\perp=0$ can then be analyzed by setting $\nabla_{\br_\perp}=0$ in Eqs.\,\,(\ref{eq:dynrho},\ref{eq:dynm}), which makes the problem effectively one-dimensional. To proceed, we employ a Newton mapping~\cite{mikhailov2012foundations,caussin2014emergent,solon2015pattern} 
to cast the characterization of the interfaces into a classical mechanics problem.

\begin{figure}[t!]
	\center
        \includegraphics[width=1\linewidth]{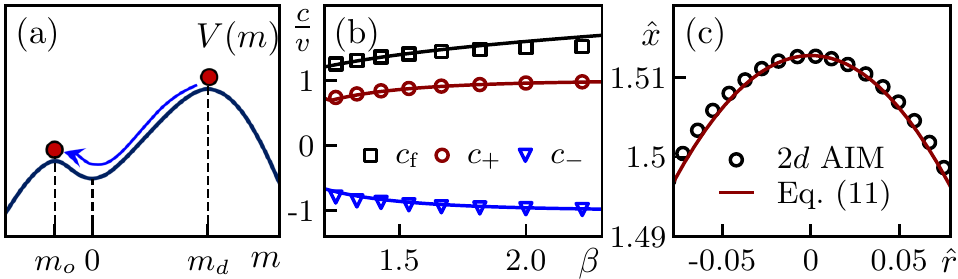}\\
    	\caption{{\bf (a)} Potential $V(m)$ entering the Newton mapping with an illustration of the heterocline corresponding to the interface between the droplet and
        the ordered phase. 
        {\bf (b)} Speeds of the 3 interfaces as a function of $\beta$.
        Solid lines correspond to values predicted by the Newton mapping. Symbols are
        measurements from microscopic simulations with 
        $(L_x, L_y) = (3000, 300)$ and $\rho_0 = 30$. Parameters: $D=v=1$. 
        {\bf (c)} Comet front profiles aligned at $\hat r=0$ from 2d AIM simulations at $t=4800$ and as
        predicted by Eq.~\eqref{eq:front-solution}.}
	\label{Fig:analytic}
\end{figure}

We first discuss the front interface between the droplet and the ordered background, whose speed we denote by $\cf$. Introducing a rescaled coordinate in the putative comoving frame,
$z=x-\cf t$, we look for stationary solutions of Eqs.\,\,(\ref{eq:dynrho},\ref{eq:dynm}), which reduce to
\begin{align}
  D_x \rho''&= v m'-\cf \rho' \label{eq:dynrho-1d}\\
  D_x m''&= v \rho'-\cf m' - F(\rho,m) \label{eq:dynm-1d}
\end{align}
where primes denote derivatives with respect to $z$. 
At leading order in a gradient expansion, Eq.~\eqref{eq:dynrho-1d} is solved by $\rho(z)\simeq \rho_\ro + v [m(z)-m_\ro]/\cf$, where $m_\ro$ is the magnetization in the ordered phase. Using this
in Eq.~\eqref{eq:dynm-1d} then leads to
\begin{equation} \label{eq:m_newton}
    D_x m'' = - \gamma(\cf) m' - \partial_m V(m,\cf)\;,
\end{equation}
where $\gamma(\cf)=v(\cf/v - v/\cf)$ and $\partial_m V(m;\cf)=F[\rho(m,\cf),m]$. Interpreting $z$ as time and $m$ as a position, Eq.~\eqref{eq:m_newton}
describes the dynamics of a fictitious particle of mass $D_x$ in a potential $V$
with a friction coefficient $\gamma$ that can be positive or negative depending on whether $\cf$ is larger or smaller than $v$, respectively.

Propagating fronts correspond to heteroclines $m(z)$ connecting extrema of $V(m)$, which is always bimodal in the ordered phase (Fig.~\ref{Fig:analytic}(a)). When $v=0$, $V(m)$ is symmetric and heteroclines exist only for
$\gamma=\cf=0$, which corresponds to static domain walls and a droplet magnetization $\md =-m_{\rm o}$, as in equilibrium. When $v\neq 0$, $\cf=0$ corresponds to a diverging friction and no static heterocline exist: fronts are thus always propagating, $V(m)$ is asymmetric so that $|m_{\rm d}|\neq |m_{\rm o}|$, and $\cf=v (m_{\rm d}-m_{\rm o})/(\rho_{\rm d}-\rho_{\rm o})$. To show that such heteroclines exist, we look for a trajectory $m(z)$ that corresponds ---without loss of generality--- to a right-going droplet. The fictitious particle then starts with a vanishing speed $m'=0$ at $m=\md$, the positive maximum of $V$, and ends---also with a vanishing speed---at $m=m_{\rm o}<0$. Equation~\eqref{eq:dynrho-1d} implies that $\md=\frac{\cf-vp_{\rm o}}{\cf+vp_{\rm o}} m_{\rm o}$,
where $p_{\rm o}=m_{\rm o}/\rho_{\rm o}\in [-1,0]$, so that $\rho_{\rm d}=\frac{\cf-vp_{\rm o}}{\cf+vp_{\rm o}} \rho_{\rm o}$.
Using the explicit expression of $F$, one finds that $\Delta V\equiv V(\md)-V(m_0)>0$
so that we need a positive friction ---and thus $\cf\geq v$--- that
dissipates exactly the energy $\Delta V$. By
continuity, a solution with $v<\cf<\infty$ always exists as shown by considering the limiting cases:
For $\cf = v$,
$\gamma(c) = 0$, the energy is conserved and the particle overshoots
to $m = -\infty$; for $\cf \to \infty$, the friction diverges
and the particle ends up trapped at the minimum $m = 0$. The value of $c_\mathrm{f}$ can then be found, e.g., by dichotomy.

In the low temperature regime considered here, Eqs.\,\,(\ref{eq:dynrho},\ref{eq:dynm}) 
predict an empty comet with $c_\pm=\mp v p_{\rm o}$. To resolve the finite density $\rho_{\rm c}$ of the disordered comet reported in Fig.~\ref{fig:droplet-micro}(a-c), one needs to follow~\cite{solon2013revisiting,martin2021fluctuation} 
and include fluctuation-induced corrections to the mean-field approximation by replacing $F$ by $F-\alpha  m/\rho$ in Eq.~\eqref{eq:dynm}, with $\alpha$ a constant. As shown in~\cite{supp}, this suffices to characterize all interfaces, at the cost of introducing the unknown constant $\alpha$. The front speeds then differ from their mean-field values by a correction of order $\rho_{\rm c}/\rho_{\rm o}$, which is negligible in the large $\rho_{\rm o}$ regime considered here. 
In Fig.~\ref{Fig:analytic}(b), we compare the mean-field predictions of our domain-wall theory for the front speeds 
to microscopic simulations in the large $\rho_{\rm o}$ limit, where mean-field is expected to work well. 
Despite the crude approximate solution of Eq.~\eqref{eq:dynrho-1d}, the agreement is very good without any fitting parameters. Importantly, our theory predicts that $c_{\rm f}>v>c_+$ so that the droplet grows in time. Similarly, the comet spreads at a speed $c_+-c_-=2v|p_{\rm o}|$. We note that a more complex structure emerges as $\beta$ decreases away from the low-temperature limit studied here. In particular, the density of the comet increases, eventually leading to its polarization. This regime and the corresponding transition will be detailled in a future publication~\cite{UsFuture}.

\if{
The other two fronts, connecting droplet to comet and comet to ordered phase, can be studied using similar techniques. However, since the comet has a low density, one needs to follow \cite{solon2013revisiting,martin2021fluctuation} and include fluctuation-induced corrections to the mean-field approximation by replacing $F$ by $F-\alpha  m/\rho$ in Eq.~\eqref{eq:dynm}, with $\alpha$ a constant. As shown in SM \cite{supp}, this then suffices to characterize all interfaces. In the large density/low temperature regime considered here, these fronts are predicted  to propagate at speeds $c_\pm \simeq \mp v p_{\rm o}$, independent of $\alpha$, so that the comet spreads at a speed $c_+-c_-=2v|p_{\rm o}|$. 
In Fig.~\ref{Fig:analytic}(c), we compare the predictions of our domain-wall theory for the speeds of the three fronts to microscopic simulations in the large $\rho_{\rm o}$ limit, where mean-field is expected to work well. Despite our crude truncation of Eq.~\eqref{eq:sum-expansion}, the agreement is very good without any fitting parameters. }\fi

We now use the mean-field Eqs.\,\,(\ref{eq:dynrho},\ref{eq:dynm}) to characterize the transverse droplet shape, which is constant at large time under the ballistic scaling. The diffusion terms in
Eqs.\,\,(\ref{eq:dynrho},\ref{eq:dynm}) then decay as $1/t$ and can be neglected. Using the radial symmetry of the droplet,  Eq.~\eqref{eq:dynrho}  reduces to
\begin{equation}
  \label{eq:rho-ball}
(\hat x \partial_{\hat x} + \hat r\partial_{\hat r}) \hat \rho = v \partial_{\hat x} \hat m\;,
\end{equation}
where $\hat{\rho}$ and $\hat{m}$ are the density and magnetization fields expressed in the reduced coordinates
$\hat{x}$ and  $\hat r$=$|\bfr_\perp/t|$.

Our domain-wall theory predicts vanishing interface widths under the ballistic scaling~\footnote{Note that Fig.~\ref{fig:droplet-micro} is thus far from this late-time regime.}. Therefore, the droplet shape is entirely characterized by the curves $\hat x_\pm(\hat r)$ and $\hat x_{\rm f}(\hat r)$ in the large-time limit. (See Fig.~\ref{fig:micro-PDE}(a) for an illustration in 2d, where $\hat r=|\hat y|$.)  The three interfaces can be characterized using a similar method. First, we center the profiles around an interface $\hat x_i(\hat r)$ by changing variables to $(\bar x,\bar r)=(\hat x-\hat x_i(\hat r), \hat r)$. Using the chain rule and integrating $\bar x$ from $-\epsilon$ to $\epsilon$ then leads to:
\begin{equation}\label{eq:interface}
[\hat x_i(\hat r)-\hat r \partial_{\hat r} \hat x_i(\hat r) ] \Delta \hat \rho = v \Delta \hat m +{\cal O}(\epsilon)\;,
\end{equation}
where $\Delta \hat \rho$ and $\Delta \hat m$ are the density and magnetization differences across the interface. In the comet, the magnetization vanishes and the density is much smaller than $\rho_{\rm o}$, so that ${\hat x}_-(\hat r)$ satisfies
$\hat x_--\hat r \partial_{\hat r} \hat x_-  = v p_{\rm o}$.
Inside the droplet, the ballistic scaling enforces $F=0$ at long times so that $\hat m(\hat x,\hat r)= -p_{\rm o} \hat \rho(\hat x,\hat r)$. For the comet-droplet interface, Eq.~\eqref{eq:interface} then leads to 
$\hat x_+(\hat r)-\hat r \partial_{\hat r} \hat x_+(\hat r)  = -v p_{\rm o}$.  These equations are readily solved by linear relations:
\begin{equation} \label{eq:vacuumintsol}
 \hat x_\pm(\hat r)=\mp v p_{\rm o} + a_\pm \hat r\;,
\end{equation}
where $a_\pm$ are finite constants. ($a_\pm=\infty$ corresponds to
bands spanning the system.) The case $a_\pm=0$ corresponds to a degenerate droplet that does not spread in the transverse directions, which is forbidden for $D_\perp\neq 0$.
Note that finite $a_\pm$ correspond to the straight interfaces shown in Figs.~\ref{fig:droplet-micro} and~\ref{fig:micro-PDE}.

Finally, we consider the interface between the droplet and the ordered phase. 
Since $\rho_{\rm d}$ diverges as $\rho_{\rm o}\to \infty$, the density and magnetization jumps require a closer inspection. Inside the droplet, using $\hat m=-p_{\rm o} \hat \rho$, 
Eq.~\eqref{eq:rho-ball} can be rewritten as
\begin{equation}
  \label{eq:rho-ball-drop}
\left[(\hat x-c_+) \partial_{\hat x} + \hat r \partial_{\hat r})\right] \hat \rho  = 0\;,
\end{equation}
where we have used $c_+\simeq - v p_{\rm o}$. This implies that $\hat \rho$ is a function of $\hat r/(\hat x-c_+)$ so that, in the large-time limit, isodensity surfaces are cones originating at $(\hat x=c_+,\hat
r=0)$. 
In particular, density and magnetization are constant along $\hat r=0$, equal to $\rho_{\rm d}$ and $m_{\rm d}=-p_{\rm o} \rho_{\rm d}$, respectively. For small $\hat r$, we thus expand the density profile as
\begin{equation}
  \label{eq:rho-expand}
  \hat \rho(\tfrac{\hat r}{\hat x-c_+})=\rho_{\rm d} [1-\tfrac k 2  (\tfrac{\hat r}{\hat x-c_+})^2]\;.
\end{equation}
To solve the interface Eq.~\eqref{eq:interface}, we first use that $\hat m= p_{\rm o} \rho_{\rm o}$ in the ordered phase and $\hat m = -p_{\rm o} \hat \rho$ in the droplet, to express $v \frac{\Delta \hat m}{\Delta \hat \rho}$ in terms of $\hat \rho$. Using Eq.~\eqref{eq:rho-expand} and the expression of $\cf$, one then gets perturbatively in $\hat r$
\begin{equation}
  \label{eq:front-solution}
  \xf=\cf-\frac{k\rho_{\rm d}}{4 v |p_{\rm o}| \rho_{\rm o}} {\hat r}^2+{\cal O}(\hat r^4).
\end{equation}
The transverse density modulation inside the droplet thus leads to the curved shape of the front interface. To compare with simulations, we first estimate the coefficient $k$ by fitting
Eq.~\eqref{eq:rho-expand} to the density field inside the droplet, and then use it to compare the prediction of
Eq.~\eqref{eq:front-solution} with the measured shape, which shows a good
agreement (Fig.~\ref{Fig:analytic}(c)).

{\it Discussion.} We have shown that the ordered phase of the active Ising model is metastable to the nucleation of minority-phase droplets. Using mean-field theory, we have revealed how the ballistic spreading of the droplet results from the selection of its domain walls and we predicted its asymptotic shape in any dimension. 
Our results imply that the stability of ordered phases with a discrete symmetry is very different in flocking and equilibrium models.

As suggested by our hydrodynamic analysis, we expect our results to be robust to microscopic details and to apply broadly to flocking models with discrete-symmetry order parameters. For instance, we verified that off-lattice versions of the AIM~\cite{martin2021fluctuation} also have a metastable ordered phase. We also predict that our results will apply to flocking models with continuous symmetries in the presence of rotational anisotropy~\cite{solon2022susceptibility}. 
While one expects mean-field results to hold better in higher dimensions, the stability of our growing droplet
solutions should be confirmed numerically beyond the $2d$ case studied here. 

From a broader perspective, our results suggest two scenarios for the flocking phase of active models, 
given the numerical results obtained on the Vicsek model in \cite{codina2022small}, 
that suggest that metastability only occurs
in a fraction of the Toner-Tu phase.
Either these results are confirmed analytically, making systems with discrete symmetries less stable than their continuous-symmetry counterparts---in opposition to the equilibrium case---, or the Toner-Tu phase 
is also generically metastable. 

\acknowledgments
DM acknowledges the support of the Center for Scientific Excellence of the Weizmann Institute of Science. 
JT thanks ANR THEMA for financial support and MSC laboratory for hospitality. YK, OG and RY acknowledge financial support from ISF (2038/21) and NSF/BSF (2022605). OG also acknowledges support from the Adams Fellowship Program of the Israeli Academy of Science and Humanities. 

\bibliographystyle{apsrev4-2}
\bibliography{biblio}

\begin{thebibliography}{32}%
\makeatletter
\providecommand \@ifxundefined [1]{%
 \@ifx{#1\undefined}
}%
\providecommand \@ifnum [1]{%
 \ifnum #1\expandafter \@firstoftwo
 \else \expandafter \@secondoftwo
 \fi
}%
\providecommand \@ifx [1]{%
 \ifx #1\expandafter \@firstoftwo
 \else \expandafter \@secondoftwo
 \fi
}%
\providecommand \natexlab [1]{#1}%
\providecommand \enquote  [1]{``#1''}%
\providecommand \bibnamefont  [1]{#1}%
\providecommand \bibfnamefont [1]{#1}%
\providecommand \citenamefont [1]{#1}%
\providecommand \href@noop [0]{\@secondoftwo}%
\providecommand \href [0]{\begingroup \@sanitize@url \@href}%
\providecommand \@href[1]{\@@startlink{#1}\@@href}%
\providecommand \@@href[1]{\endgroup#1\@@endlink}%
\providecommand \@sanitize@url [0]{\catcode `\\12\catcode `\$12\catcode
  `\&12\catcode `\#12\catcode `\^12\catcode `\_12\catcode `\%12\relax}%
\providecommand \@@startlink[1]{}%
\providecommand \@@endlink[0]{}%
\providecommand \url  [0]{\begingroup\@sanitize@url \@url }%
\providecommand \@url [1]{\endgroup\@href {#1}{\urlprefix }}%
\providecommand \urlprefix  [0]{URL }%
\providecommand \Eprint [0]{\href }%
\providecommand \doibase [0]{https://doi.org/}%
\providecommand \selectlanguage [0]{\@gobble}%
\providecommand \bibinfo  [0]{\@secondoftwo}%
\providecommand \bibfield  [0]{\@secondoftwo}%
\providecommand \translation [1]{[#1]}%
\providecommand \BibitemOpen [0]{}%
\providecommand \bibitemStop [0]{}%
\providecommand \bibitemNoStop [0]{.\EOS\space}%
\providecommand \EOS [0]{\spacefactor3000\relax}%
\providecommand \BibitemShut  [1]{\csname bibitem#1\endcsname}%
\let\auto@bib@innerbib\@empty
\bibitem [{\citenamefont {Mermin}\ and\ \citenamefont
  {Wagner}(1966)}]{mermin1966absence}%
  \BibitemOpen
  \bibfield  {author} {\bibinfo {author} {\bibfnamefont {N.~D.}\ \bibnamefont
  {Mermin}}\ and\ \bibinfo {author} {\bibfnamefont {H.}~\bibnamefont
  {Wagner}},\ }\href@noop {} {\bibfield  {journal} {\bibinfo  {journal}
  {Physical Review Letters}\ }\textbf {\bibinfo {volume} {17}},\ \bibinfo
  {pages} {1133} (\bibinfo {year} {1966})}\BibitemShut {NoStop}%
\bibitem [{\citenamefont {Kardar}(2007)}]{kardar2007statistical}%
  \BibitemOpen
  \bibfield  {author} {\bibinfo {author} {\bibfnamefont {M.}~\bibnamefont
  {Kardar}},\ }\href@noop {} {\emph {\bibinfo {title} {Statistical physics of
  fields}}}\ (\bibinfo  {publisher} {Cambridge University Press},\ \bibinfo
  {year} {2007})\BibitemShut {NoStop}%
\bibitem [{\citenamefont {Vicsek}\ \emph {et~al.}(1995)\citenamefont {Vicsek},
  \citenamefont {Czir{\'o}k}, \citenamefont {Ben-Jacob}, \citenamefont
  {Cohen},\ and\ \citenamefont {Shochet}}]{vicsek1995novel}%
  \BibitemOpen
  \bibfield  {author} {\bibinfo {author} {\bibfnamefont {T.}~\bibnamefont
  {Vicsek}}, \bibinfo {author} {\bibfnamefont {A.}~\bibnamefont {Czir{\'o}k}},
  \bibinfo {author} {\bibfnamefont {E.}~\bibnamefont {Ben-Jacob}}, \bibinfo
  {author} {\bibfnamefont {I.}~\bibnamefont {Cohen}},\ and\ \bibinfo {author}
  {\bibfnamefont {O.}~\bibnamefont {Shochet}},\ }\href@noop {} {\bibfield
  {journal} {\bibinfo  {journal} {Physical Review Letters}\ }\textbf {\bibinfo
  {volume} {75}},\ \bibinfo {pages} {1226} (\bibinfo {year}
  {1995})}\BibitemShut {NoStop}%
\bibitem [{\citenamefont {Toner}\ and\ \citenamefont
  {Tu}(1995)}]{toner1995long}%
  \BibitemOpen
  \bibfield  {author} {\bibinfo {author} {\bibfnamefont {J.}~\bibnamefont
  {Toner}}\ and\ \bibinfo {author} {\bibfnamefont {Y.}~\bibnamefont {Tu}},\
  }\href@noop {} {\bibfield  {journal} {\bibinfo  {journal} {Physical Review
  Letters}\ }\textbf {\bibinfo {volume} {75}},\ \bibinfo {pages} {4326}
  (\bibinfo {year} {1995})}\BibitemShut {NoStop}%
\bibitem [{\citenamefont {Toner}\ \emph {et~al.}(2005)\citenamefont {Toner},
  \citenamefont {Tu},\ and\ \citenamefont
  {Ramaswamy}}]{toner2005hydrodynamics}%
  \BibitemOpen
  \bibfield  {author} {\bibinfo {author} {\bibfnamefont {J.}~\bibnamefont
  {Toner}}, \bibinfo {author} {\bibfnamefont {Y.}~\bibnamefont {Tu}},\ and\
  \bibinfo {author} {\bibfnamefont {S.}~\bibnamefont {Ramaswamy}},\ }\href@noop
  {} {\bibfield  {journal} {\bibinfo  {journal} {Annals of Physics}\ }\textbf
  {\bibinfo {volume} {318}},\ \bibinfo {pages} {170} (\bibinfo {year}
  {2005})}\BibitemShut {NoStop}%
\bibitem [{\citenamefont {Toner}(2012)}]{toner2012reanalysis}%
  \BibitemOpen
  \bibfield  {author} {\bibinfo {author} {\bibfnamefont {J.}~\bibnamefont
  {Toner}},\ }\href@noop {} {\bibfield  {journal} {\bibinfo  {journal}
  {Physical Review E}\ }\textbf {\bibinfo {volume} {86}},\ \bibinfo {pages}
  {031918} (\bibinfo {year} {2012})}\BibitemShut {NoStop}%
\bibitem [{\citenamefont {Chat{\'e}}\ \emph {et~al.}(2008)\citenamefont
  {Chat{\'e}}, \citenamefont {Ginelli}, \citenamefont {Gr{\'e}goire},\ and\
  \citenamefont {Raynaud}}]{chate2008collective}%
  \BibitemOpen
  \bibfield  {author} {\bibinfo {author} {\bibfnamefont {H.}~\bibnamefont
  {Chat{\'e}}}, \bibinfo {author} {\bibfnamefont {F.}~\bibnamefont {Ginelli}},
  \bibinfo {author} {\bibfnamefont {G.}~\bibnamefont {Gr{\'e}goire}},\ and\
  \bibinfo {author} {\bibfnamefont {F.}~\bibnamefont {Raynaud}},\ }\href@noop
  {} {\bibfield  {journal} {\bibinfo  {journal} {Physical Review E}\ }\textbf
  {\bibinfo {volume} {77}},\ \bibinfo {pages} {046113} (\bibinfo {year}
  {2008})}\BibitemShut {NoStop}%
\bibitem [{\citenamefont {Mahault}\ \emph {et~al.}(2019)\citenamefont
  {Mahault}, \citenamefont {Ginelli},\ and\ \citenamefont
  {Chat{\'e}}}]{mahault2019quantitative}%
  \BibitemOpen
  \bibfield  {author} {\bibinfo {author} {\bibfnamefont {B.}~\bibnamefont
  {Mahault}}, \bibinfo {author} {\bibfnamefont {F.}~\bibnamefont {Ginelli}},\
  and\ \bibinfo {author} {\bibfnamefont {H.}~\bibnamefont {Chat{\'e}}},\
  }\href@noop {} {\bibfield  {journal} {\bibinfo  {journal} {Physical Review
  Letters}\ }\textbf {\bibinfo {volume} {123}},\ \bibinfo {pages} {218001}
  (\bibinfo {year} {2019})}\BibitemShut {NoStop}%
\bibitem [{\citenamefont {Codina}\ \emph {et~al.}(2022)\citenamefont {Codina},
  \citenamefont {Mahault}, \citenamefont {Chat{\'e}}, \citenamefont {Dobnikar},
  \citenamefont {Pagonabarraga},\ and\ \citenamefont {Shi}}]{codina2022small}%
  \BibitemOpen
  \bibfield  {author} {\bibinfo {author} {\bibfnamefont {J.}~\bibnamefont
  {Codina}}, \bibinfo {author} {\bibfnamefont {B.}~\bibnamefont {Mahault}},
  \bibinfo {author} {\bibfnamefont {H.}~\bibnamefont {Chat{\'e}}}, \bibinfo
  {author} {\bibfnamefont {J.}~\bibnamefont {Dobnikar}}, \bibinfo {author}
  {\bibfnamefont {I.}~\bibnamefont {Pagonabarraga}},\ and\ \bibinfo {author}
  {\bibfnamefont {X.-q.}\ \bibnamefont {Shi}},\ }\href@noop {} {\bibfield
  {journal} {\bibinfo  {journal} {Physical Review Letters}\ }\textbf {\bibinfo
  {volume} {128}},\ \bibinfo {pages} {218001} (\bibinfo {year}
  {2022})}\BibitemShut {NoStop}%
\bibitem [{\citenamefont {Besse}\ \emph {et~al.}(2022)\citenamefont {Besse},
  \citenamefont {Chat{\'e}},\ and\ \citenamefont
  {Solon}}]{besse2022metastability}%
  \BibitemOpen
  \bibfield  {author} {\bibinfo {author} {\bibfnamefont {M.}~\bibnamefont
  {Besse}}, \bibinfo {author} {\bibfnamefont {H.}~\bibnamefont {Chat{\'e}}},\
  and\ \bibinfo {author} {\bibfnamefont {A.}~\bibnamefont {Solon}},\
  }\href@noop {} {\bibfield  {journal} {\bibinfo  {journal} {Physical Review
  Letters}\ }\textbf {\bibinfo {volume} {129}},\ \bibinfo {pages} {268003}
  (\bibinfo {year} {2022})}\BibitemShut {NoStop}%
\bibitem [{\citenamefont {Solon}\ \emph {et~al.}(2022)\citenamefont {Solon},
  \citenamefont {Chat{\'e}}, \citenamefont {Toner},\ and\ \citenamefont
  {Tailleur}}]{solon2022susceptibility}%
  \BibitemOpen
  \bibfield  {author} {\bibinfo {author} {\bibfnamefont {A.}~\bibnamefont
  {Solon}}, \bibinfo {author} {\bibfnamefont {H.}~\bibnamefont {Chat{\'e}}},
  \bibinfo {author} {\bibfnamefont {J.}~\bibnamefont {Toner}},\ and\ \bibinfo
  {author} {\bibfnamefont {J.}~\bibnamefont {Tailleur}},\ }\href@noop {}
  {\bibfield  {journal} {\bibinfo  {journal} {Physical Review Letters}\
  }\textbf {\bibinfo {volume} {128}},\ \bibinfo {pages} {208004} (\bibinfo
  {year} {2022})}\BibitemShut {NoStop}%
\bibitem [{\citenamefont {Solon}\ and\ \citenamefont
  {Tailleur}(2013)}]{solon2013revisiting}%
  \BibitemOpen
  \bibfield  {author} {\bibinfo {author} {\bibfnamefont {A.}~\bibnamefont
  {Solon}}\ and\ \bibinfo {author} {\bibfnamefont {J.}~\bibnamefont
  {Tailleur}},\ }\href@noop {} {\bibfield  {journal} {\bibinfo  {journal}
  {Physical Review Letters}\ }\textbf {\bibinfo {volume} {111}},\ \bibinfo
  {pages} {078101} (\bibinfo {year} {2013})}\BibitemShut {NoStop}%
\bibitem [{\citenamefont {O'Loan}\ and\ \citenamefont
  {Evans}(1999)}]{o1999alternating}%
  \BibitemOpen
  \bibfield  {author} {\bibinfo {author} {\bibfnamefont {O.}~\bibnamefont
  {O'Loan}}\ and\ \bibinfo {author} {\bibfnamefont {M.}~\bibnamefont {Evans}},\
  }\href@noop {} {\bibfield  {journal} {\bibinfo  {journal} {Journal of Physics
  A: Mathematical and General}\ }\textbf {\bibinfo {volume} {32}},\ \bibinfo
  {pages} {L99} (\bibinfo {year} {1999})}\BibitemShut {NoStop}%
\bibitem [{\citenamefont {Benvegnen}\ \emph {et~al.}(2022)\citenamefont
  {Benvegnen}, \citenamefont {Chat{\'e}}, \citenamefont {Krapivsky},
  \citenamefont {Tailleur},\ and\ \citenamefont
  {Solon}}]{benvegnen2022flocking}%
  \BibitemOpen
  \bibfield  {author} {\bibinfo {author} {\bibfnamefont {B.}~\bibnamefont
  {Benvegnen}}, \bibinfo {author} {\bibfnamefont {H.}~\bibnamefont
  {Chat{\'e}}}, \bibinfo {author} {\bibfnamefont {P.~L.}\ \bibnamefont
  {Krapivsky}}, \bibinfo {author} {\bibfnamefont {J.}~\bibnamefont
  {Tailleur}},\ and\ \bibinfo {author} {\bibfnamefont {A.}~\bibnamefont
  {Solon}},\ }\href@noop {} {\bibfield  {journal} {\bibinfo  {journal}
  {Physical Review E}\ }\textbf {\bibinfo {volume} {106}},\ \bibinfo {pages}
  {054608} (\bibinfo {year} {2022})}\BibitemShut {NoStop}%
\bibitem [{\citenamefont {Domb}\ and\ \citenamefont
  {Green}(1972)}]{griffiths1972phase}%
  \BibitemOpen
  \bibinfo {editor} {\bibfnamefont {C.}~\bibnamefont {Domb}}\ and\ \bibinfo
  {editor} {\bibfnamefont {M.~S.}\ \bibnamefont {Green}},\ eds.,\ \href@noop {}
  {\emph {\bibinfo {title} {'Rigorous Results and Theorems', by R.B. Griffiths
  in Phase Transitions and Critical Phenomena}}},\ Vol.~\bibinfo {volume} {1}\
  (\bibinfo  {publisher} {Academic Press, London},\ \bibinfo {year}
  {1972})\BibitemShut {NoStop}%
\bibitem [{\citenamefont {Peierls}(1991)}]{peierls1991more}%
  \BibitemOpen
  \bibfield  {author} {\bibinfo {author} {\bibfnamefont {R.}~\bibnamefont
  {Peierls}},\ }\href@noop {} {\emph {\bibinfo {title} {More surprises in
  theoretical physics}}}\ (\bibinfo  {publisher} {Princeton University Press},\
  \bibinfo {year} {1991})\BibitemShut {NoStop}%
\bibitem [{\citenamefont {Solon}\ and\ \citenamefont
  {Tailleur}(2015)}]{solon2015flocking}%
  \BibitemOpen
  \bibfield  {author} {\bibinfo {author} {\bibfnamefont {A.~P.}\ \bibnamefont
  {Solon}}\ and\ \bibinfo {author} {\bibfnamefont {J.}~\bibnamefont
  {Tailleur}},\ }\href@noop {} {\bibfield  {journal} {\bibinfo  {journal}
  {Physical Review E}\ }\textbf {\bibinfo {volume} {92}},\ \bibinfo {pages}
  {042119} (\bibinfo {year} {2015})}\BibitemShut {NoStop}%
\bibitem [{\citenamefont {Sakaguchi}\ and\ \citenamefont
  {Ishibashi}(2019)}]{sakaguchi2019flip}%
  \BibitemOpen
  \bibfield  {author} {\bibinfo {author} {\bibfnamefont {H.}~\bibnamefont
  {Sakaguchi}}\ and\ \bibinfo {author} {\bibfnamefont {K.}~\bibnamefont
  {Ishibashi}},\ }\href@noop {} {\bibfield  {journal} {\bibinfo  {journal}
  {Physical Review E}\ }\textbf {\bibinfo {volume} {100}},\ \bibinfo {pages}
  {052113} (\bibinfo {year} {2019})}\BibitemShut {NoStop}%
\bibitem [{\citenamefont {Yu}\ and\ \citenamefont {Tu}(2022)}]{yu2022energy}%
  \BibitemOpen
  \bibfield  {author} {\bibinfo {author} {\bibfnamefont {Q.}~\bibnamefont
  {Yu}}\ and\ \bibinfo {author} {\bibfnamefont {Y.}~\bibnamefont {Tu}},\
  }\href@noop {} {\bibfield  {journal} {\bibinfo  {journal} {Physical Review
  Letters}\ }\textbf {\bibinfo {volume} {129}},\ \bibinfo {pages} {278001}
  (\bibinfo {year} {2022})}\BibitemShut {NoStop}%
\bibitem [{\citenamefont {Mishra}\ and\ \citenamefont
  {Mishra}(2022)}]{mishra2022active}%
  \BibitemOpen
  \bibfield  {author} {\bibinfo {author} {\bibfnamefont {P.~K.}\ \bibnamefont
  {Mishra}}\ and\ \bibinfo {author} {\bibfnamefont {S.}~\bibnamefont
  {Mishra}},\ }\href@noop {} {\bibfield  {journal} {\bibinfo  {journal}
  {Physics of Fluids}\ }\textbf {\bibinfo {volume} {34}},\ \bibinfo {pages}
  {057110} (\bibinfo {year} {2022})}\BibitemShut {NoStop}%
\bibitem [{\citenamefont {Kourbane-Houssene}\ \emph {et~al.}(2018)\citenamefont
  {Kourbane-Houssene}, \citenamefont {Erignoux}, \citenamefont {Bodineau},\
  and\ \citenamefont {Tailleur}}]{kourbane2018exact}%
  \BibitemOpen
  \bibfield  {author} {\bibinfo {author} {\bibfnamefont {M.}~\bibnamefont
  {Kourbane-Houssene}}, \bibinfo {author} {\bibfnamefont {C.}~\bibnamefont
  {Erignoux}}, \bibinfo {author} {\bibfnamefont {T.}~\bibnamefont {Bodineau}},\
  and\ \bibinfo {author} {\bibfnamefont {J.}~\bibnamefont {Tailleur}},\
  }\href@noop {} {\bibfield  {journal} {\bibinfo  {journal} {Physical review
  letters}\ }\textbf {\bibinfo {volume} {120}},\ \bibinfo {pages} {268003}
  (\bibinfo {year} {2018})}\BibitemShut {NoStop}%
\bibitem [{\citenamefont {Chat{\'e}}(2020)}]{chate2020dry}%
  \BibitemOpen
  \bibfield  {author} {\bibinfo {author} {\bibfnamefont {H.}~\bibnamefont
  {Chat{\'e}}},\ }\href@noop {} {\bibfield  {journal} {\bibinfo  {journal}
  {Annual Review of Condensed Matter Physics}\ }\textbf {\bibinfo {volume}
  {11}},\ \bibinfo {pages} {189} (\bibinfo {year} {2020})}\BibitemShut
  {NoStop}%
\bibitem [{sup()}]{supp}%
  \BibitemOpen
  \href@noop {} {}\bibinfo {note} {See Supplemental Material [url], which
  includes theoretical and numerical details, as well as Refs. XXX}\BibitemShut
  {NoStop}%
\bibitem [{Note1()}]{Note1}%
  \BibitemOpen
  \bibinfo {note} {Keeping $r$ fixed and increasing $\rho _{\protect \rm d}^0$
  leads to similar results, see Fig.\protect \,\protect \,S2}\BibitemShut
  {NoStop}%
\bibitem [{\citenamefont {Bray}(2002)}]{bray2002theory}%
  \BibitemOpen
  \bibfield  {author} {\bibinfo {author} {\bibfnamefont {A.~J.}\ \bibnamefont
  {Bray}},\ }\href@noop {} {\bibfield  {journal} {\bibinfo  {journal} {Advances
  in Physics}\ }\textbf {\bibinfo {volume} {51}},\ \bibinfo {pages} {481}
  (\bibinfo {year} {2002})}\BibitemShut {NoStop}%
\bibitem [{\citenamefont {Krapivsky}\ \emph {et~al.}(2010)\citenamefont
  {Krapivsky}, \citenamefont {Redner},\ and\ \citenamefont
  {Ben-Naim}}]{krapivsky2010kinetic}%
  \BibitemOpen
  \bibfield  {author} {\bibinfo {author} {\bibfnamefont {P.~L.}\ \bibnamefont
  {Krapivsky}}, \bibinfo {author} {\bibfnamefont {S.}~\bibnamefont {Redner}},\
  and\ \bibinfo {author} {\bibfnamefont {E.}~\bibnamefont {Ben-Naim}},\
  }\href@noop {} {\emph {\bibinfo {title} {A kinetic view of statistical
  physics}}}\ (\bibinfo  {publisher} {Cambridge University Press},\ \bibinfo
  {year} {2010})\BibitemShut {NoStop}%
\bibitem [{\citenamefont {Mikhailov}(2012)}]{mikhailov2012foundations}%
  \BibitemOpen
  \bibfield  {author} {\bibinfo {author} {\bibfnamefont {A.~S.}\ \bibnamefont
  {Mikhailov}},\ }\href@noop {} {\emph {\bibinfo {title} {Foundations of
  synergetics I: Distributed active systems}}},\ Vol.~\bibinfo {volume} {51}\
  (\bibinfo  {publisher} {Springer Science \& Business Media},\ \bibinfo {year}
  {2012})\BibitemShut {NoStop}%
\bibitem [{\citenamefont {Caussin}\ \emph {et~al.}(2014)\citenamefont
  {Caussin}, \citenamefont {Solon}, \citenamefont {Peshkov}, \citenamefont
  {Chat{\'e}}, \citenamefont {Dauxois}, \citenamefont {Tailleur}, \citenamefont
  {Vitelli},\ and\ \citenamefont {Bartolo}}]{caussin2014emergent}%
  \BibitemOpen
  \bibfield  {author} {\bibinfo {author} {\bibfnamefont {J.-B.}\ \bibnamefont
  {Caussin}}, \bibinfo {author} {\bibfnamefont {A.}~\bibnamefont {Solon}},
  \bibinfo {author} {\bibfnamefont {A.}~\bibnamefont {Peshkov}}, \bibinfo
  {author} {\bibfnamefont {H.}~\bibnamefont {Chat{\'e}}}, \bibinfo {author}
  {\bibfnamefont {T.}~\bibnamefont {Dauxois}}, \bibinfo {author} {\bibfnamefont
  {J.}~\bibnamefont {Tailleur}}, \bibinfo {author} {\bibfnamefont
  {V.}~\bibnamefont {Vitelli}},\ and\ \bibinfo {author} {\bibfnamefont
  {D.}~\bibnamefont {Bartolo}},\ }\href@noop {} {\bibfield  {journal} {\bibinfo
   {journal} {Physical Review Letters}\ }\textbf {\bibinfo {volume} {112}},\
  \bibinfo {pages} {148102} (\bibinfo {year} {2014})}\BibitemShut {NoStop}%
\bibitem [{\citenamefont {Solon}\ \emph {et~al.}(2015)\citenamefont {Solon},
  \citenamefont {Caussin}, \citenamefont {Bartolo}, \citenamefont {Chat{\'e}},\
  and\ \citenamefont {Tailleur}}]{solon2015pattern}%
  \BibitemOpen
  \bibfield  {author} {\bibinfo {author} {\bibfnamefont {A.}~\bibnamefont
  {Solon}}, \bibinfo {author} {\bibfnamefont {J.-B.}\ \bibnamefont {Caussin}},
  \bibinfo {author} {\bibfnamefont {D.}~\bibnamefont {Bartolo}}, \bibinfo
  {author} {\bibfnamefont {H.}~\bibnamefont {Chat{\'e}}},\ and\ \bibinfo
  {author} {\bibfnamefont {J.}~\bibnamefont {Tailleur}},\ }\href@noop {}
  {\bibfield  {journal} {\bibinfo  {journal} {Physical review. E, Statistical,
  nonlinear, and soft matter physics}\ }\textbf {\bibinfo {volume} {92 6}},\
  \bibinfo {pages} {062111} (\bibinfo {year} {2015})}\BibitemShut {NoStop}%
\bibitem [{\citenamefont {Martin}\ \emph {et~al.}(2021)\citenamefont {Martin},
  \citenamefont {Chat{\'e}}, \citenamefont {Nardini}, \citenamefont {Solon},
  \citenamefont {Tailleur},\ and\ \citenamefont
  {Van~Wijland}}]{martin2021fluctuation}%
  \BibitemOpen
  \bibfield  {author} {\bibinfo {author} {\bibfnamefont {D.}~\bibnamefont
  {Martin}}, \bibinfo {author} {\bibfnamefont {H.}~\bibnamefont {Chat{\'e}}},
  \bibinfo {author} {\bibfnamefont {C.}~\bibnamefont {Nardini}}, \bibinfo
  {author} {\bibfnamefont {A.}~\bibnamefont {Solon}}, \bibinfo {author}
  {\bibfnamefont {J.}~\bibnamefont {Tailleur}},\ and\ \bibinfo {author}
  {\bibfnamefont {F.}~\bibnamefont {Van~Wijland}},\ }\href@noop {} {\bibfield
  {journal} {\bibinfo  {journal} {Physical Review Letters}\ }\textbf {\bibinfo
  {volume} {126}},\ \bibinfo {pages} {148001} (\bibinfo {year}
  {2021})}\BibitemShut {NoStop}%
\bibitem [{UsF()}]{UsFuture}%
  \BibitemOpen
  \href@noop {} {}\bibinfo {note} {Benvegnen, B. and Granek, O. and Ro, S. and
  Yaacoby, R. and Chaté, H. and Kafri, Y. and Mukamel, D. and Solon, A. and
  Tailleur, J., In Preparation}\BibitemShut {NoStop}%
\bibitem [{Note2()}]{Note2}%
  \BibitemOpen
  \bibinfo {note} {Note that Fig.~\ref {fig:droplet-micro} is thus far from
  this late-time regime.}\BibitemShut {Stop}%
\end{thebibliography}%

\end{document}